# Coulomb oscillations of a quantum antidot formed by an airbridged pillar gate in the integer and fractional quantum Hall regime

Tokuro Hata[1*], Hiroki Mitani[1], Hidetaka Uchiyama[1], Takafumi Akiho[2], Koji Muraki[2] and Toshimasa Fujisawa[1]

[1]*Department of Physics, Tokyo Institute of Technology, 2-12-1 Ookayama, Meguro-ku, Tokyo 152-8551, Japan*
[2]*NTT Basic Research Laboratories, 3-1 Morinosato Wakamiya, Atsugi-shi, Kanagawa 243-0198, Japan*
*E-mail: hata@phys.titech.ac.jp
† Tokyo Institute of Technology merged with Tokyo Medical and Dental University to form Institute of Science Tokyo (Science Tokyo) on October 1, 2024.

Quantum antidots (QAD) are attractive for manipulating quasiparticles in quantum Hall (QH) systems. Here, we form a QAD in the integer and fractional QH regimes at nominal Landau-level filling factor $\bar{\nu}$ = 2, 1, and 2/3 using a submicron pillar gate with an airbridge connection. After confirming the required conditions for a fully depleted QAD, we analyze the observed Coulomb oscillations in terms of the area of the QAD and the effective charge for the oscillation period in an identical gate voltage range. The area at $\bar{\nu}$ = 2/3 is significantly smaller than that at $\bar{\nu}$ = 2 and 1, in qualitative agreement with the previous report. By assuming a constant gate capacitance, the effective charge at $\bar{\nu}$ = 2/3 is about 2/3 of that at $\bar{\nu}$ = 2 and 1. The QAD device can be used to capture and emit charges in the unit of $2e/3$.

## 1. Introduction

The fractional quantum Hall (QH) effect can be seen by applying a perpendicular high magnetic field to a high-mobility two-dimensional electron gas (2DEG). The quasiparticles associated with fractional QH systems have unique fractional charges ($e/3$, $e/5$, etc. with elementary charge $e$), as experimentally demonstrated by shot noise measurements [1,2]. Furthermore, such fractional-charge quasiparticles have recently been verified to exhibit anyon statistics by several schemes [3-8]. For instance, the exchange phase associated with a fractional charge bound to an impurity state has been identified with a fractional-charge Fabry–Pérot interferometer [3,4]. The exchange phase is also confirmed with the two-particle partition noise under collisions of fractional charges stochastically generated from two quantum point contact [5-8]. For more advanced experiments with fractional-charge anyons, it is desirable to develop novel functional devices that can control fractional charges at will, such as a quantum antidot (QAD) [9-16]. An experimental scheme for braiding fractional-charge anyons has been proposed with several QADs in a scalable way [17]. Such QADs can be formed by locally depleting electrons in the integer and fractional QH regimes [18-24]. Particularly in the fractional regime, the charge changes in the unit of the effective charge, which is a fractional multiple of the elementary charge. Therefore, fractional charges can be emitted or captured by a QAD prepared in a fractional QH system. Experimentally, QADs can be defined, for example with multilayered gates [24] and etched grooves on GaAs/AlGaAs heterostructures [18] and graphene layers [25,26]. We have been working on fabricating submicron QADs using a submicron pillar gate connected by an airbridge structure. In the integer QH regime, we have successfully formed a single QAD with a submicron diameter [27,28] but also multiple tunnel-coupled QADs [29]. Even in the fractional QH regime, such a device shows clear Coulomb oscillations [30], and thus systematic analyses of the effective charge are highly desirable.

In this study, we investigate the fundamental characteristics, such as effective charge, of a single QAD defined by a submicron pillar gate with an airbridge connection in the integer and fractional QH regimes. After describing the condition for forming a fully depleted QAD surrounded by a well-defined QH state with a nominal Landau-level filling factor $\bar{\nu}$, we analyze Coulomb oscillations obtained at various magnetic fields and gate voltages for $\bar{\nu}$ = 2, 1, and 2/3. The period in the magnetic field can be used to determine the area of the QAD and thus depletion spreading $d$ around the pillar gate; $d \simeq 0.1$ μm for $\bar{\nu}$ = 2/3 is shorter than $d \simeq 0.3$ μm for $\bar{\nu}$ = 1. The period in the gate voltage should be proportional to the effective charge $e^*$ under a constant gate capacitance; $e^*$ for $\bar{\nu}$ = 2/3 is about 2/3 of elementary charge $e$ for integer $\bar{\nu}$ (= 1 and 2). The device is promising for manipulating fractional charges.

## 2. QAD device with airbridge gate

The QAD sample used in this study was fabricated in a standard GaAs/AlGaAs heterostructure with 2DEG located at $t$ = 100 nm below the surface. The schematic plan view of the device in Fig. 1(a) shows the airbridge structure with a cylindrical pillar of diameter $D$ = 300 nm, an airbridge of length $L$ = 3 μm and width $W$ = 300 nm, and a bridge support. Two side gates S$_L$ and S$_R$ are placed with spacing $s$ = 700 nm to the pillar. These gates with total thickness $H$ = 300 nm were fabricated by depositing Ti (30 nm) and Au (270 nm) films after patterning a triple layer resist by using electron-beam lithography [27-29]. As shown by the cross-section in Fig. 1(b), the pillar is electrically connected through the airbridge with bridge height $h$ = 150 nm to the bridge support on the left. The bridge structure of a control sample is confirmed with the scanning electron micrograph of Fig. 1(c) taken from a 45-degree oblique direction.

The edge channels can be formed, as shown by the red lines with arrows in Fig. 1(a), by applying perpendicular magnetic field $B$ and appropriate gate voltages $V_{gB}$ on the pillar through the bridge, $V_{gSL}$ on S$_L$, and $V_{gSR}$ on S$_R$. $V_{gSL}$ and $V_{gSR}$ are applied so that the left and right channels are coupled to the QAD channel. Here, the Landau-level filling factor $\nu_A$ in the airbridge region is different from $\nu_U$ in the ungated region, because finite $V_{gB}$ changes the electron density under the airbridge. To form a well-defined QAD, the 2DEGs in the ungated region and under the airbridge should share the same QH state with the nominal filling factor $\bar{\nu}$, which is an integer or fractional value to describe the quantized Hall conductance of the QH states, as we describe below. Electrons under the pillar should be fully depleted with local filling factor $\nu_P$ (= 0), and thus the QAD is defined as a closed channel, as shown by the red circle around the pillar. Electrons under the pillar are depleted in the unit of effective charge $e^*$. The objective of this study is to estimate this $e^*$ for some integer and fractional QH states.

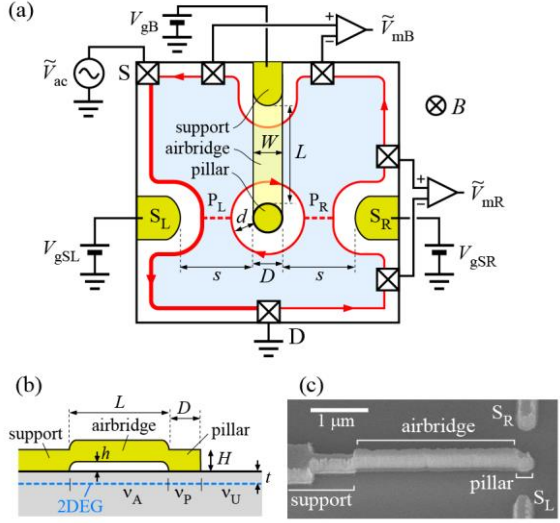

Fig. 1. (a) Schematic of the QAD device with the airbridge structure, which is made of the pillar, airbridge, and support. The localized edge channel(s) around the pillar (the red circle) constitutes the QAD. Two side gates $S_L$ and $S_R$ are used to investigate the transport through the QAD and tunneling paths $P_L$ and $P_R$. The ac voltage $\tilde{V}_{ac}$ is applied, and the voltage drop $\tilde{V}_{mR}$ is measured. (b) Schematic cross-section of the airbridge structure. The 2DEG is characterized by Landau-level filling factors $\nu_A$ under the airbridge, $\nu_P$ under the pillar, and $\nu_U$ in the ungated region. (c) Scanning electron micrograph of a control device with $L = 2$ μm and $D = W = 0.3$ μm.

As shown by the red dashed lines in Fig. 1(a), transport through the QAD takes place via tunneling paths $P_L$ to the left channel and $P_R$ to the right channel. We apply ac voltage $\tilde{V}_{ac}$ = 30 μV at 37 Hz to the source (S), and thus the chemical potential of the channel (the thick red line) from the source to the grounded drain (D) is modulated with $e\tilde{V}_{ac}$. The current can be obtained by measuring the voltage drop $\tilde{V}_{mR}$ between the voltage probes across $S_R$ by using a lock-in amplifier. Namely, current $\tilde{I}_{mR} = g\tilde{V}_{mR}$ through the QAD can be obtained with $g = \bar{\nu}e^2/h$. When the 2DEG under the airbridge is not in a QH state, a leakage current flows to induce a finite voltage drop $\tilde{V}_{mB}$ across the bridge gate. We measure $\tilde{V}_{mB}$ to identify the QH state under the airbridge as well as in the ungated region.

The device has a back gate to tune the electron density $n$ of the 2DEG, and we fixed at $n \sim 1.6 \times 10^{15}$ m$^{-2}$ with the back-gate voltage of about -100 V to change $\bar{\nu}$ = 2, 1, and 2/3 within the maximum field (12 T) of the magnet. All measurements were performed at ~ 100 mK.

## 3. QAD condition

To compare the characteristics of QADs for different QH states, the QADs should be formed with identical gate voltages in the selected $B$ ranges for each QH state. We describe the required conditions in this section. To define a fully depleted QAD with $\nu_P = 0$, $V_{gB}$ has to be sufficiently negative below the pinch-off voltage $V_{pG} \sim$ -0.3 V for the gated region. Here, $V_{pG} \sim$ -0.3 V is obtained from a separate pinch-off measurement at zero magnetic field (not shown). The negative $V_{gB}$ reduces $\nu_A$ under the airbridge region. Therefore, a QH state in the ungated and under the airbridge regions appears in the QH-U and QH-A regions, respectively, in the schematic $B$ - $V_{gB}$ plane of Fig. 2(a). As a result, the well-defined QAD surrounded by the same QH state can be found in the yellow triangular region surrounded by the red dashed line. Here, the pinch-off voltage $V_{pA}$ under the airbridge should be much more negative than $V_{pG}$ with the expected ratio $V_{pA}/V_{pG} \sim (1 + \varepsilon_r h/t) = 20.5$ for the dielectric constant $\varepsilon_r \sim 13$ of GaAs.

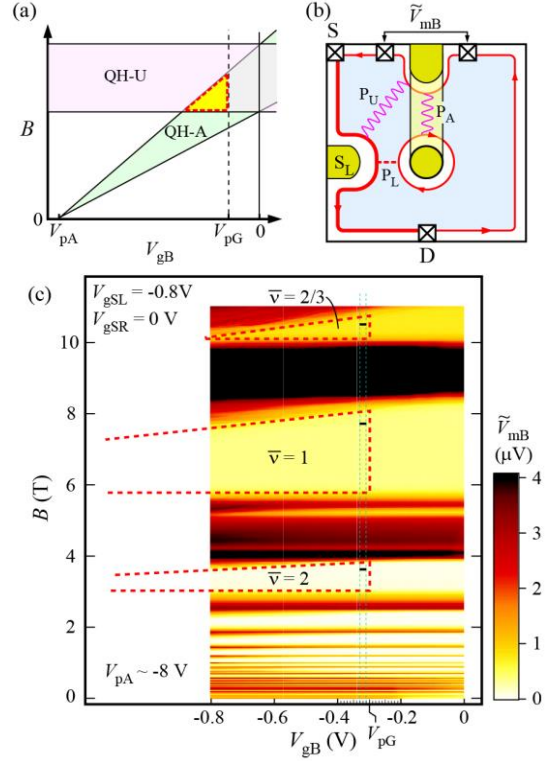

Fig. 2. (a) The condition (the yellow triangular region) for defining a fully depleted QAD ($V_{gB} < V_{pG}$) surrounded by an identical QH state in the ungated region (QH-U) and under the airbridge (QH-A). (b) Schematic setup for testing the bulk transport through diffusive paths $P_U$ in the ungated region and $P_A$ under the airbridge. (c) Color plot of $\tilde{V}_{mB}$ as a function of $B$ and $V_{gB}$ under $\tilde{V}_{ac}$ = 30 μV. The conditions for a fully depleted QAD is shown by the red dashed lines for $\bar{\nu}$ = 2/3, 1, and 2.

To find the above condition experimentally, $\tilde{V}_{mB}$ across the bridge is investigated as a function of $B$ and $V_{gB}$ with $V_{gSL} < V_{pG}$ and $V_{gSR} = 0$ V, as shown in Fig. 2(b), where the possible QAD is coupled to the left channel. Then, $\tilde{V}_{mB}$ is sensitive to the bulk transport in the ungated region and under the airbridge region, as shown by the wavy lines $P_U$ and $P_A$.

Figure 2(c) shows the color-scale plot of $\tilde{V}_{mB}$ observed at $V_{gSL}$ = -0.8 V and $V_{gSR}$ = 0 V. The condition for no bulk transport in the ungated and under the airbridge regions appears as $\tilde{V}_{mB} \sim$ 0, inside the red dashed lines for $\bar{\nu}$ = 2, 1, and 2/3. Here, $V_{pG}$ = -0.3 V is assumed for the maximum $V_{gB}$ (the vertical red dashed lines). By extrapolating the slanting red dashed lines to the horizontal axis, we obtain large negative $V_{pA} = -8$ V in our device. $V_{pA}/V_{pG} = -8/-0.3 \sim 26.6$ is comparable to the expected value above. While a QAD can be formed anywhere in the designated regions, we focus on QADs in the same range of $V_{gB} = -330$ – $-310$ mV (between the cyan dotted lines) at $B \sim 3.6$ T for $\bar{\nu} = 2$, $\sim 7.7$ T for $\bar{\nu} = 1$, and $\sim 10.5$ T for $\bar{\nu} = 2/3$ (marked by the small black regions).

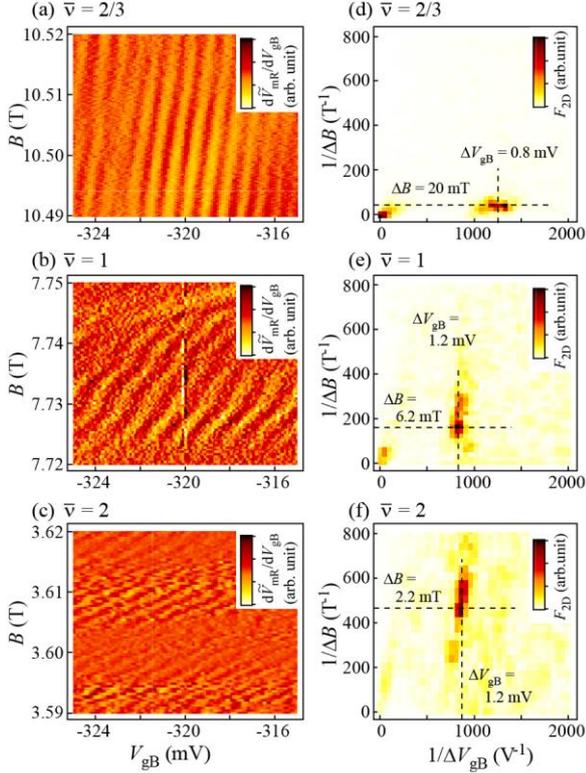

Fig. 3. (a-c) Coulomb oscillations observed in the derivative $d\tilde{V}_{mR}/dV_{gB}$ for (a) $\bar{\nu} = 2/3$, (b) 1, and (c) 2. (d-f) Two-dimensional Fourier spectrum $F_{2D}$ for (d) $\bar{\nu} = 2/3$, (e) 1, and (f) 2, from which oscillation periods $\Delta V_{gB}$ and $\Delta B$ are obtained.

Table. I. Analysis of the Coulomb-oscillation period $\Delta B$ around the magnetic field $B$ for $\bar{\nu}$. The area $S$ of the QAD is deduced by assuming $N$ interacting edge channels. The depletion length $d$ is estimated for the pillar diameter $D = 0.3$ μm.

| $\bar{\nu}$ | $B$ [T] | $\Delta B$ [mT] | $N$ | $S$ [(μm)$^2$] | $d$ [μm] |
|---|---|---|---|---|---|
| 2 | ~3.6 | 2.2 | 2 | 0.94 | 0.40 |
| 1 | ~7.7 | 6.2 | 1 | 0.66 | 0.31 |
| 2/3 | ~10.5 | 20 | 1 | 0.21 | 0.11 |

Table. II. Analysis of the Coulomb-oscillation period $\Delta V_{gB}$ around the magnetic field $B$ for $\bar{\nu}$. The effective charge $e^*$ is obtained by assuming a constant gate capacitance and $e^* = e$ at $\bar{\nu} = 1$.

| $\bar{\nu}$ | $B$ [T] | $\Delta V_{gB}$ [mV] | $\Delta V_{gB}/\Delta V_{gB,1}$ ($\equiv e^*/e$) |
|---|---|---|---|
| 2 | ~3.6 | 1.2 | 1 |
| 1 | ~7.7 | 1.2 | $\equiv 1$ |
| 2/3 | ~10.5 | 0.8 | 0.67 |

### 4. Coulomb oscillations

Coulomb oscillations of the QAD can be observed by changing the magnetic field and the gate voltage. The period in the magnetic field is primarily determined by the Aharonov-Bohm (AB) effect, which quantizes the area $S$ of the electronic orbits as specified by the relation $BS = m\phi_0$ with the magnetic flux quanta $\phi_0 = h/e$ and an integer $m$. Therefore, the oscillation period $\Delta B = \phi_0/S$ in $B$ should be given by $S$. When the QAD is formed by electrostatically coupled $N$ edge channels with comparable area $S$, the oscillation frequency is multiplied by $N$ with reduced period $\Delta B = \phi_0/NS$. Here, $N$ depends on charge dynamics between the edge channels, and a previous report suggests $N = \bar{\nu}$ for integer $\bar{\nu}$ and $N = 1$ for $\bar{\nu} = 2/3$ [24].

On the other hand, the oscillation period $\Delta V_{gB}$ in the gate voltage $V_{gB}$ reflects the effective charge $e^*$ with the relation $e^* = C\Delta V_{gB}$ for capacitance $C$ between the gate and the QAD. For a constant $C$, $\Delta V_{gB}$ is proportional to $e^*$. We use this relation to measure $e^*$ for fractional QH state at $\bar{\nu} = 2/3$. For this purpose, we fixed the sweep range of $V_{gB}$ to maintain the bare electron-density distribution up to the occupation in Landau levels.

The Coulomb oscillations are measured as a function of $B$ and $V_{gB}$ at $V_{gSL} = -0.9$ V and $V_{gSR} = -1.2$ V. Its derivative $d\tilde{V}_{mR}/dV_{gB}$ is shown for better visibility in Fig. 3(a) for $\bar{\nu} = 2/3$, Fig. 3(b) for $\bar{\nu} = 1$, and Fig. 3(c) for $\bar{\nu} = 2$, where the identical span (30 mT) of the $B$ ranges and the identical $V_{gB}$ range are used. Because the oscillation patterns are not purely periodic but are wiggled possibly due to impurity states of the 2DEG, we employed two-dimensional (2D) Fourier transform as shown in Fig. 3(d) for $\bar{\nu} = 2/3$, Fig. 3(e) for $\bar{\nu} = 1$, and Fig. 3(f) for $\bar{\nu} = 2$. While the 2D Fourier spectrum $F_{2D}$ is broadened in a certain direction, representative peak positions are shown by the dashed lines with the periods $\Delta B$ and $\Delta V_{gB}$. These periods are analyzed with the model described above, as summarized in Table I for $\Delta B$ and Table II for $\Delta V_{gB}$.

For $\Delta B$, the area $S$ of the QAD is obtained by assuming $N$ interacting edge channels. The depletion length $d$ around the pillar with diameter $D = 0.3$ μm is estimated by assuming a circular QAD orbit. Our result shows $d_{2/3} \simeq 0.1$ μm for $\bar{\nu} = 2/3$ is shorter than $d_1 \simeq 0.3$ μm for $\bar{\nu} = 1$. This shrink ratio $d_{2/3}/d_1 \sim 0.33$ is somewhat smaller than that ($d_{2/3}/d_1 \sim 0.6$) reported with larger QADs with lithographic diameter of $1 - 2$ μm [24]. The difference might be related to the detailed edge structure of the smaller device.

For $\Delta V_{gB}$, the effective charge $e^*$ is obtained by assuming a constant gate capacitance and $e^* = e$ at $\bar{\nu} = 1$. Namely, $e^*/e$ is given by the ratio $\Delta V_{gB}/\Delta V_{gB,1}$ with the period $\Delta V_{gB,1}$ for $\bar{\nu} = 1$. The obtained $e^*/e \simeq 1$ at $\bar{\nu} = 2$ and $e^*/e \simeq 2/3$ at $\bar{\nu} = 2/3$ is consistent with the previous study [24]. This implies that the charge trapped in the QAD at $\bar{\nu} = 2/3$ changes by $e^* \simeq 2e/3$ for each oscillation period. This can be used to manipulate fractional charges. Note that the constant gate capacitance for the present 2DEG wafer has been confirmed experimentally by measuring the plasmon velocity, which is inversely proportional to the channel capacitance [31]. This supports the validity of the analysis.

### 5. Summary

We have investigated Coulomb oscillations of a QAD with a pillar gate in the integer and fractional QH regimes. The required condition to form a fully depleted QAD has been estimated by measuring the bulk transport in the ungated region and under the airbridge. We also have verified that the gate-voltage period at $\bar{\nu} = 2/3$ is almost 2/3 of that at $\bar{\nu} = 1$ and 2, which implies that the charge trapped in the QAD changes in the unit of effective charge $e^*$ close to $2e/3$ at $\bar{\nu} = 2/3$. The capability of manipulating the fractional charge with our QAD is promising for realizing anyon braiding and novel quantum electronics by integrating multiple QADs.

## Acknowledgments

This work was supported by JSPS KAKENHI Grant Numbers (JP19H05603, JP24K16993, JP24H00827, and JP24K00552, and partially conducted at Nanofab in the Tokyo Institute of Technology supported by "Advanced Research Infrastructure for Materials and Nanotechnology (ARIM)" in Japan, and Materials Analysis Division, Open Facility Center in Tokyo Institute of Technology.